\documentstyle[emulateapj]{article}
\newcommand{\mns}{m_{\rm ns}}
\newcommand{\mrg}{M_{\rm RG}}
\newcommand{\rrg}{R_{\rm RG}}
\newcommand{\msun}{M_{\odot}}
\newcommand{\rsun}{R_{\odot}}

\newcommand{\ttt}{\times}

\slugcomment{Astrophysical Journal, Vol 521, 502-507, August 20, 1999}
\begin{document}

\title{The formation and merger of compact objects
in central engine of active galactic nuclei and quasars:
gamma-ray burst and gravitational radiation}

\author {K.S. Cheng}
\affil{Physics Department, The University of Hong Kong, Hong Kong, P.R. China}
\author{Jian-Min Wang}
\affil{Astronomy Department, Beijing Normal University, Beijing 100875,
and Astronomy Department, Nanjing University, Nanjing 210008, and
Chinese Acadmey of Sciences - Peking University Joint Beijing
Astrophysical Center (CAS-PKU.BAC), Beijing 100871, P.R. China}

\begin{abstract}
The production rate of compact objects, i.e. neutron stars (NS) 
and black holes (BH), in active galactic nuclei (AGN) and quasars (QSO), 
where the frequent supernova explosion is used to explain the high 
metallicity, is very high due to the interaction between the accretion 
disk and main sequence stars in the nucleus of the quasar. The compact 
object-red giant star (RG) binaries can be easily formed due to the large
captured cross-section of the red giant stars. 
The (NS/BH, NS/BH) binary can be formed after the supernova
explosion of the (NS/BH, RG) binary. Intense transient gamma-ray emission
(gamma-ray burst) and gravitational radiation can result from the merger 
of these two compact objects. Collision between helium core (Hc) of RG and 
black hole may also take place and  may also result in long duration 
gamma-ray bursts but no gravitational waves. We estimate that the merger 
rate of (NS/BH, NS/BH) binaries and (Hc, BH) is 
proportional to the metal abundance 
$\left({\rm \frac{NV}{CIV}}\right)$ and can be as 
high as 10$^{-3}\left({\rm \frac{NV}{CIV}}/0.01\right)$ per year per AGN/QSO.

\keywords{quasar: emission line -- star: neutron star/black hole --
gamma-ray: burst -- gravitational radiation} 

\end{abstract}

\section{Introduction}

The basic physical scenario of active galactic nuclei and quasars is
generally believed to be a supermassive black hole surrounded by accretion disk
which releases the observed huge body of energy (Rees 1984). However
in this model one prominent problem is the fuel. Syer, Clark, \& Rees (1991)
investigate the possibility of accretion disk capturing stars in dense 
star cluster of nucleus (within 1pc) in order to remove the accretion
fuel problem. This interesting interaction has been employed to explain
the high metallicity phenomena in active galactic nuclei and quasars by
Artymowicz, Lin \& Wampler (1993), Zurek et al (1994), especially for
broad absorption line (BAL) quasars with extremely high metallicity 
(sometimes several hundred times that of solar abundance, Shields 1996).
It is interesting to note that this process also result in the formation 
and coalescence of compact objects in the central 
energy house of active galactic nuclei and quasars which may be 
sources of frequent $\gamma$-ray and gravitational wave bursts.

Gravitational wave could soon be detected directly even emitted from a 
source located at a cosmological distance because several 
ground-based gravitational wave experiments including  
the Laser Interferometer Gravitational-Wave Observatory 
(LIGO) (Abramovici et al. 1992), VIRGO, TAMA300 and GEO600 are 
under construction. The best understood of all gravitational-wave 
sources are coalescing, compact binaries composed of neutron stars 
and black holes. These NS/NS, NS/BH and BH/BH binaries may 
give information about the equation of state (EOS) of nuclear 
matter at high densities (Shibata, Nakamura \& Oohara 1992, 1993; 
Rasio \& Shapiro 1994; Zhuge, Centrella \& McMillan 1994; Davies et al. 
1994; Ruffert, Janka \& Sch\"afer 1996; Cheng \& Dai 1998). 

In addition the observations of afterglow of Gamma-ray Burst (GRB) 
lend strong supports to the
cosmological origin of GRB (Paczy\'nski 1986, Mao \& Paczy\'nski 1992,
Meegan et al 1992, Piran 1992) (see a concise review by Greiner
1998 for the afterglow, and Piran 1998).
Despite the effort of searching more than 20 years to look for the 
counterparts of GRB at 
other wave bands, no known population of objects in the Universe 
have been identified with any GRB sources (e.g.
Fenimore et al. 1993, Fishman \& Meegan 1995, Band \& Hartmann 1998).
Two main reasons to cause this situation. First, the position accuracy
of GRB detected by BASTE is so low that a large number of known objects
are  located in its error box. Secondly, the burst duration is so short 
that it is very difficult to use other means to reduce the error box. 
Although the BeppoSAX has observed a dozen of GRBs' afterglows, one of 
these sources can reduce its error box down to $\sim$ 
one arc second. But still no confirmed counterparts are identified,
even for the well-observed, GBR970228 and GRB970508. On the
other hand, the observed isotropic and inhomogeneous distribution of 
GRBs detected by the BASTE instrument on board the Compton Gamma-Ray 
Observatory (CGRO) is best explained by a source population at 
cosmological distances.

Most cosmological GRB models involve compact objects. For examples, 
(1) neutron star - neutron star/black hole merger
(e.g. Paczy\'nski 1986; Eichler et al 1989;
Dermer 1992; Narayan, Piran \& Shemi 1991; Mao \& Paczy\'nski 1992); 
(2) collision between black hole and Helium core or white dwarf by
Fryer \& Woosley (1998), Fryer et al (1998);
(3)the phase transition from neutron star to strange star (Cheng \& Dai 
1996); (4) the transition of normal nuclear matter to matter with pion 
condensation in neutron star (e.g. Haensel et al 1990, 
Muto et al 1993); (5) rapidly rotating neutron star with extremely large
magnetic field (Usov 1992, Duncan \& Thompson 1992);
(6) a failed supernova type Ib (Woosley 1993), or microquasar (Paczy\'nski 
1998) etc. Therefore, if we want 
to find the hosts of GRB sources, it is logical to look for cosmological 
sources with a lot of compact objects. From the stellar evolution point of 
view, the progenitors of compact objects, e.g. neutron stars, are massive 
stars (approximately $5\sim 20 M_{\odot}$). 
>From the observations of afterglows of GRBs it has been deduced by 
Paczy\'nski (1998) that GRB may associate with the star formation region.
We note that most of the masses of the massive stars, consisting of 
a lot of heavy elements, will be ejected during the 
supernova explosion and a metal-rich region will be formed. 
Naturally, galaxies with high abundance of 
compact object should be metal-rich. Therefore it is logical to look for 
cosmological sources with high metal abundance. In order to observe the 
sources located at cosmological distance, these sources must be luminous. 
It should be noted that most AGNs/QSOs are metal-rich (Artymowicz, Lin \&
Wampler 1993,
Hamann et al 1997). This may be the real reason why GRBs may be related 
to AGNs and QSOs as reported by Schartel, Andernach \& Greiner (1996),
and Burenin et al (1998) who claim 
there are possible association of $\gamma$-ray burst with radio quiet
quasar, and with active galactic nuclei, respectively. However
the physical scenario of this association remains open if it is true,
at least, some of GRB are related with AGNs/QSOs.
It has been suggested that
GRB may be produced in a star formation region which leads to 
expectation of the burst rate of GRB proportional to the formation 
rate of star in the cosmological galaxies (Paczy\'nski 1998). Because 
the lifetime of AGN/QSO is rather short
(generally less than $10^8$ yr), the possible host objects of GRB are 
mostly ones related with galaxies. Alternatively here we suggest that 
the interaction between accretion disk and stars in dense cluster may lead
to the formation of a large number compact objects in the central 
engine of active galactic
nuclei. The formation and interactions among the compact objects in the
nuclei might associate with GRB. Therefore active galactic nuclei are
one kind of its host objects if GRBs are indeed associated with compact 
objects.

\section{The Model}

The mechanism for the enhancement of 
metal-rich elements has been suggested by a way of evolution of
main sequence stars, which are captured by the accretion disk
surrounding the supermassive black hole in the center of quasar
(for a review cf. Lin \& Papaloizou 1996). The captured stars would accrete 
material from the disk and increase their masses up to 50 solar masses in 
less than $10^4$ years. These massive stars evolve rapidly toward the 
supernova stage
and the ejecta of the supernovae provide enough heavy element to produce
the features suggested by observations. 
The metal abundance expressed in terms of the ratio
between NV and CIV is related to the total number of
captured main sequence stars $N_{\rm MS}$ during the active phase of quasar 
as follows (Artymowicz, Lin \& Wampler 1993; Zurek et al 1994)
\begin{equation}
\frac{{\rm NV}}{{\rm CIV}}\approx 0.01 
		       \left(\frac{N_{\rm MS}}{1.0\times 10^5}\right).
\end{equation}
However this mechanism
results in other implications, such as formation of neutron stars/black holes
in the vicinity of the accretion disk.
As we will argue in the subsequent section, 
it exists an efficient way to form compact object - compact object 
(i.e. NS/BH, NS/BH) binaries via the formation of 
compact object - red giant star (NS/BH, RG) binaries.
The (NS/BH, NS/BH) binary will be formed after the explosion of the red 
giant star. In such a scheme the merger rate of (NS/BH, NS/BH) binary 
approximately equals to the capture rate of main sequence star, therefore 
it is proportional to the metal abundance. If the merger of (NS/BH, NS/BH) 
binary or Helium core of RG and black hole is one of the possible mechanisms 
of GRBs, then active galactic nuclei and quasars should associate with GRBs. 
In fact the coalescence of two compact objects also produces intense 
transient gravitational waves which should be detected by LIGO/VIRGO in 
1Gpc radius.

\subsection{Evolution of the captured stars in the accretion disk}

We assume that the nucleus of the quasar consists of a massive black hole 
surrounded by an accretion disk, which are enclosed by a star cluster. 
The structure of disk is described by the standard model in which
the local height and the surface density of the accretion disk are given by 
(e.g. Frank, et al 1992)
\begin{equation}
\left(\frac{h}{R}\right)=1.6\times 10^{-3}
      \alpha^{-1/10}\dot{m}^{3/20}M_8^{-1/10}r_{\rm d}^{1/8},
\end{equation}
and
\begin{equation}
\Sigma=1.7\times 10^7 \alpha^{-4/5} \dot{m}^{7/10} M_8^{1/5}r_{\rm d}^{-3/4}
       ({\rm g~cm^{-2}}),
\end{equation}
respectively, and the inward drift velocity due to the outward 
transportation of the angular momentum in disk reads
\begin{equation}
v_{\rm r}=4.1 \times 10^4 \alpha^{4/5}\dot{m}^{3/10}M_8^{-1/5}r_{\rm d}^{-1/4}
	  ({\rm cm~s^{-1}}),
\end{equation}
or the time scale of inward drift is
\begin{equation}
\tau_{\rm r}=4.0\times 10^7 \alpha^{-4/5}\dot{m}^{-3/10}M_8^{6/5}
	     \left(\frac{r_{\rm d}}{10^5}\right)^{5/4}({\rm yrs}),
\end{equation}
where $\dot{m}$ is dimensionless accretion rate scaled with the Eddington limit,
$\alpha$ is the
viscosity coefficient, $r_{\rm d}$ = $\frac{R}{R_s}$ is the radius of disk 
normalized by Schwartzchild radius 
$R_s=\frac{GM_{\rm BH}}{c^2}$, and $M_8$ denotes 
the mass of central black hole ($M_{\rm BH}$) in units of $10^8M_{\odot}$.

The main sequence stars in the star cluster will be captured by the disk and 
the number of the captured main sequence stars in the annular $R-R+dR$ is 
estimated as (Artymowicz, Lin \& Wampler 1993),
\begin{eqnarray}                                                         
dN&=&m_*G^2(\Sigma RdR)\frac{32\pi^{1/2} C_d \nu_0 \Delta T}
                            {\sigma_0^3} \nonumber\\
  &            \approx& 0.45\left(\frac{m_*}{M_{\odot}}\right)
               \left(\frac{\nu_0}{10^7{\rm pc^{-3}}}\right)
	       \left(\frac{\Delta T}{10^8{\rm yr}}\right)
	       \left(\frac{\sigma_0}{300{\rm kms^{-1}}}\right)^{-3}\nonumber\\
  &        &\alpha^{-4/5}\dot{m}^{7/10}M_8^{11/5}r_{\rm d}^{1/4} dr_{\rm d}
\end{eqnarray}
where $\Delta T$ is the active period of quasar, $\nu_0$ is the number
density of star cluster at 1 pc in the range of about $10^7$/pc$^3$
(Blandford 1991), $\sigma_0$ is the dispersion velocity of star in
the cluster, and $C_d\approx 6$. Integrating over the disk, there are  
$\sim 10^5$ main sequence stars captured by disk in the active phase of 
quasar or the captured rate is $\sim 10^{-3} yr^{-1}$. 
The captured stars will
coorotate with the medium in the disk (Syer et al 1991), and 
accrete gas from the disk with Bondi
rate until a gap appears in disk with condition that the star's Roche
radius exceeds the disk scale-height $h$ (Lin \& Papaloizou, 1996). The
timescale of Bondi accretion is
\begin{equation}
\tau_a=1.40\times 10^3\alpha^{2/5}\dot{m}^{-1/10}M_8^{2/5}
       \left(\frac{r_{\rm d}}{10^5}\right)^{3/4}
       \left(\frac{m}{M_{\odot}}\right)^{-1} ({\rm yrs}),
\end{equation}
and the maxium accreted mass of the captured star is limited by 
\begin{equation}
\left(\frac{m}{M_{\odot}}\right)=17.7 \alpha^{-3/10} 
		    \dot{m}^{9/20} M_8^{7/10}
		    \left(\frac{r_{\rm d}}{10^5}\right)^{3/8}.
\end{equation}
This condition is also
coincided with that the Bondi radius ($r_{\rm B}=Gm_*/c_{\rm s}^2$)
does not exceed the local height $h$. With the typical parameters 
of the quasars and $\alpha$ = 1, the maximum accreted mass of the 
captured stars is in the range of 10 $\sim$ 20 $M_{\odot}$ (but the 
maximum mass of the captured stars could be much higher than 20 
$M_{\odot}$ if $\alpha$ is much less than unity).
The star in this mass range will evolve off their
main sequence quickly and likely become  neutron stars plus small 
fraction of black holes (see e.g. Shapiro \& Teukolsky 1983). For more 
massive stars (m $\ge$ 20 $M_{\odot}$) they could become neutron stars 
or black holes (Timmes, Woosley \& Weaver 1996).
The simple formula of the evolution time scale for the main
sequence star is (Meurs \& van den Heuvel 1989)
\begin{equation}
\tau_{\rm e}=10^{a_1}\left(\frac{m}{M_{\odot}}\right)^{a_2},
\end{equation}
where the index $a_1$ and $a_2$ are tabulated in their table 3.
For interested cases, we have $a_1=9.3$, $a_2=-2$ for
$3.8\le m/M_{\odot}\le 12$, and $a_1=8.2$, $a_2=-1$ for star
larger than $12M_{\odot}$.
Substituting parameters for lower mass case into Eq.9, we obtain:
\begin{equation}
\tau_{\rm e}=9.0\times 10^6 \alpha^{3/10}\dot{m}^{-9/20}M_8^{-7/10}
	     \left(\frac{r_{\rm d}}{10^5}\right)^{-3/8}~~({\rm yrs}).
\end{equation}
The higher mass case will evolve faster than that given in Eq.(10). It has been estimated that the main sequence stars start to be
captured by the disk at the disk radius $\sim$1 pc 
(Artymowicz, Lin \&  Wampler 1993)
but the exact value of this radius is not crucial in our model.
The more important radius $R_c$ is before which the captured stars 
finish the evolution of supernova stage. This radius must be larger
than that of the tidal radius $R_{\rm t}$ = 
$\left(\frac{M_{\rm BH}}{m_*}\right)^{1/3}r_*$, otherwise the captured 
stars will disrupted by the central massive black hole. 
This radius can be estimated by equating the radial inward drifting
time scale Eq.(5) and the evolution time scale 
Eq.(11) and is given by
\begin{equation}
r_{\rm c}=\frac{R_{\rm c}}{R_{\rm s}}
	 =4.0\times 10^4 \alpha^{44/65}\dot{m}^{-6/65}M_8^{-76/65}.
\end{equation}
It appears that this radius is always larger than that of $R_t$ for typical
quasar parameters.  

\subsection{Formation of binaries}

In the early stage, almost all of the captured stars can evolve to
neutron stars/black holes with mass $\mns$
which will be ejected from the disk after supernova
explosion. But these neutron stars/black holes can only shoot up to a 
scale height
\begin{equation}
h_{\rm ns}=\frac{v_{\infty}^2 R_{\rm c}^2}{G M_{\rm BH}},
\end{equation}
where $v_{\infty}\sim$ 300 km/s is the neutron star kick velocity produced by 
supernova explosion. Taking the typical parameters, $h_{\rm ns}\sim$ 
$10^{16}$ cm. The compact stars will be oscillating up and down 
crossing  the disk. Artymowicz, Lin \& Wampler (1993) also note the 
possibility of retrapping of compact stars by disk, but the following 
case is more interesting.
Before evolving to compact star, the captured star will go through
the red giant phase with radius ($\rrg$) and mass $\mrg$. 
During the close encounter with a neutron star, the
red giant star would undergo substantial tidal deformation 
at the cost of a part of the relative kinetic energy of the orbit.
Such a tidal process can eventually dissipate
the total positive energy of the initial unbound orbit via oscillations
and heating, and a (NS/BH, RG) binary system will be created. One might argue
that the velocity dispersion of NS/BH is much larger than the escape velocity
of RG at its surface (about 70km/s). However, it is interesting to note 
the following situation.  The surface density of RG
$\Sigma_{\rm RG}\sim M_{\rm RG}/R_{\rm RG}^2\sim 10^7$ g/cm$^2$ is
much larger than that of disk $\sim 10^3$ g/cm$^2$ at 0.1pc.
This leads to an efficient
dissipation of NS/BH kinetic energy in the process of encountering RG's
envelope. The probability
of NS/BH encountering the dense envelope of RG can be easily estimated
\begin{equation}
f=\frac{S_{\rm RG}}{S_{\rm disk}}\sim 10^{-2},
\end{equation}
where $S_{\rm RG}$ is the total surface area of all RG, and $S_{\rm disk}$ 
is the area of disk capturing main sequence star. The drag force acting on
NS/BH by the dense envelope of RG reads (Artymowicz et al 1993)
\begin{equation}
F_d=\frac{4\pi G^2m_{\rm ns}^2\rho C_d}{v_{\infty}^2}.
\end{equation}
After $N$ times of encountering the disk, the kinetic energy of NS/BH will 
be reduced
significantly so that NS/BH can be captured by RG through tidal process,
\begin{eqnarray}
N&=&\frac{v_{\infty}^4}{8\pi fG^2 m_{\rm ns}C_d \Sigma_{\rm RG}}\nonumber\\
 &\sim & 10^4\left(\frac{f}{0.01}\right)^{-1}
	 \left(\frac{\Sigma_{\rm RG}}{10^7{\rm g/cm^2}}\right)^{-1}
	 \left(\frac{v_{\infty}}{300{\rm km/s}}\right)^4,
\end{eqnarray}
and the capture time scale of NS/BH by the RG envelope is about
\begin{equation}
\tau_{\rm ns, RG}=N\frac{h_{\rm ns}}{v_{\infty}}\sim 10^5{\rm yr},
\end{equation}
which is slightly shorter than the life time of RG, therefore a large 
fraction of RG can be captured by the compact objects in the vicinity 
of the disk to form (NS/BH, RG) binaries if we take into account the 
interaction between NS/BH and the envelope of RG. Let us consider the 
tidal capture rate in more detail. The distance $D$ for the tidal capture to
form a (NS/BH, RG) binary can be estimated as
(Bhattacharya \& van den Heuvel 1991)
\begin{equation}
D \approx 1.75\rrg \left(\frac{\mns}{\mrg}\frac{\mns+\mrg}{\msun}
	  \frac{\rsun}{\rrg}\right)^{\frac{1}{6}}
	  \left(\frac{50{\rm Km/s}}{v_{\rm ns}}\right)^{\frac{1}{3}},
\end{equation}
where $v_{\rm ns}$ is the reduced velocity of the compact star. The cross section of this
capture process is given by
\begin{equation}
\sigma \approx \pi D^2\left[1+\frac{2G(\mns+\mrg)}{v^2_{\rm ns}}\right],
\end{equation}
and is $\sim 3 \times 10^{28} cm^2$ for typical parameters.
The formation rate of (NS/BH, RG) binary
thus can be written as
\begin{equation}
\frac{dN_{{\rm NR}}}{dt}=
       \left(\frac{N_{\rm ns}}{V_{\rm ns}}\right)
       \left(\frac{N_{\rm RG}}{V_{\rm RG}}\right)v_{\rm ns}\sigma V_{\rm RG} 
       =k N_{\rm ns}N_{\rm RG},
\end{equation}
where $N_{\rm ns}$ and $V_{\rm ns}\sim h_{ns} R_c^2$ 
($N_{\rm RG}$ and $V_{\rm RG}$ )
are the number and occupied volume of compact stars (red giant stars)
respectively, and constant 
$k=v_{\rm ns}\sigma/V_{\rm ns}$.
It should be noted that $N_{\rm ns}$ and $N_{\rm RG}$ are
time-dependent because the presence of (NS/BH, RG) binary formation results
from the tidal interaction between compact stars and red giant stars. 
The time-dependent numbers of compact
stars and red giant stars obey the following equations
\begin{equation}
\frac{dN_{\rm RG}}{dt}=\Gamma-kN_{\rm ns}N_{\rm RG},
\end{equation}
\begin{equation}
\frac{dN_{\rm ns}}{dt}=\frac{N_{\rm RG}}{\tau_{\rm RG}}
		       -kN_{\rm ns}N_{\rm RG},
\end{equation}
where $\Gamma$ is the capture rate of main sequence stars by disk, and 
$\tau_{\rm RG}$ represents the evolution time scale from red giant star to
NS/BH (about a few times of $10^5$ years).  
These non-linear equations can be analytically resolved
in some interesting stages. 

(1) When $t\le \frac{1}{\Gamma}$, the production of red giant stars
dominates, we have 
\begin{equation}
N_{\rm RG} \approx \Gamma t,~~~ {\rm and}~~~
N_{\rm ns} \approx \left(\frac{\Gamma}{\tau_{\rm RG}}\right)t^2. 
\end{equation}

2) In steady state, the numbers of compact stars and red giant stars 
reach constants, i.e. 
\begin{equation}
N_{\rm RG}(\infty)=\tau_{\rm RG}\Gamma,~~~~ {\rm and}~~~
N_{\rm ns}(\infty)=\frac{1}{k\tau_{\rm RG}}.
\end{equation}
It is clear that the (NS, RG) binary formation rate is a constant, i.e.
$\frac{dN_{{\rm NR}}}{dt}=\Gamma$.

\subsection{Coalescence of (NS/BH, NS/BH) and (BH, Hc) binaries}

The evolution of massive binaries leads to two possibilities: (1) Merging
of Helium core of RG and black hole, (2) (NS/BH, NS/BH) binary formations 
and merging.  The first case occurs for very close binaries in which the 
black hole companion  encounters with the Helium core before the secondary 
evolves into compact object. This process is recently suggested as a GRB 
mechanism by Fryer \& Woosley (1998).  The second corresponds to the case 
which the secondary evolves into compact star faster than that of spiral-in 
process. Then the (NS/BH, NS/BH) binaries are formed and eventually merge to 
emit intense transient gamma-rays and graviational waves. The estimations 
of above processes are given below.

The captured compact star may spiral-in the red giant star with the
core mass $M_{\rm core}$ and envelope mass $M_{\rm en}$. Many detail
calculations have been done (e.g. Bodenheimer \& Taam 1984;
Taam \& Bodenheimer 1989, Taamr  et al 1997). 
A rough estimation of final separation of the binary after
spiral-in can be made by comparing the binding energy of envelope
with the difference in total energy of the binary before and after spiral-in
(Webbink et al 1983, Verbunt 1993) 
\begin{equation}
\frac{D_f}{D_i}=\frac{M_{\rm core}}{\mrg}
		\left(1+\frac{2D_i}{\alpha_{*} \lambda R_L}
		\frac{M_{\rm en}}{\mns}\right)^{-1},
\end{equation}
where $D_i$, and $D_f$ are the initial and final separation  
respectively, $R_L$ is the Roche radius, which is given by
(Eggleton 1983)
\begin{equation}
\frac{R_L}{D_i}=\frac{0.49}{0.6+q^{-2/3}\ln (1+q^{1/3})}, 
\end{equation}
where $q$ is the mass ratio of (NS/BH, RG) binary system,
$\alpha_*$ is the efficiency with wich released binary energy is
used to lift the envelope; $\lambda$  a factor depending on the mass
distribution in the envelope of RG. Their product is often taken to be 0.20.
The core mass of red giant star for $\mrg \ge 7\msun$ is given by (Iben 1993),
\begin{equation}
\frac{M_{\rm core}}{\msun}=0.058 \left(\frac{\mrg}{\msun}\right)^{1.57}.
\end{equation}
And the radius of core is
given by (Paczy\'nski \& Kozlowski 1972; Pols \& Marinus 1994; Wijers 1998)
\begin{equation}
\frac{R_{\rm core}}{\rsun}=
	     0.22\left(\frac{M_{\rm core}}{\msun}\right)^{0.6}.
\end{equation}
If $D_f$ is smaller than the sum 
$R_{\rm core}+R_{\rm ns}\approx R_{\rm core}$, after spiral-in the compact 
star merges with the core of the red giant star. We have calculated
some typical cases listed in Table 1. It can be found that the RG
massive than 18$M_{\odot}$ will merger with the Helium core before
they form a (NS/BH, NS/BH) binary. Even the companion star is a neutron
star, it may accrete efficiently material from the red giant to become
a black hole during the spiral-in process(Chevalier 1996;
Fryer \& Woosley 1998). Therefore the merger of (NS/BH, Hc) binary can
become $\gamma$-ray burst as described by Fryer \& Woosley (1998).

   There are two possible ways to destroy the formed (NS/RG, RG) binaries:
   1) the combining gravitation of central black hole and the dense
   star cluster, 2) the third encounter from the cluster. It has been
   shown by Saslaw (1985) that the chance of the later way is rather
   small. We can now estimate the action of the first way. If the
   tidal force from the gravitation of central massive black hole and
   dense star cluster exceeds the interacting force between the two
   components, then the (NS/BH, RG) will be disrupted by the tidal 
   force. The tidal force reads
   \begin{equation}
   F_{\rm tid}=\frac{G(M_{\rm BH}+M_{\rm SC})(m_{\rm c}+M_{\rm RG})}{R^2}
               \frac{D_i}{R},
   \end{equation}
   where $M_{\rm SC}$ is the mass of dense star cluster, and $m_{\rm c}$ 
   is the mass of the compact object. Then the ratio between the tidal
   force and the gravitational force ($F_B$)
   between the two components is
   \begin{equation}
   \frac{F_{\rm tid}}{F_{\rm B}}=\frac{M_{\rm BH}+M_{\rm SC}}{M_{\rm RG}}
        \frac{M_{\rm RG}+m_{\rm c}}{m_{\rm c}}\left(\frac{D_i}{R}\right)^3
        \approx 10^{-4},
   \end{equation}
   which means that the formed binaries can survive in its surroundings.
   On the other hand, consequently the spiral-in process takes place in 
   the binary very fast. The timescale
   of this process can be roughly estimated under the assumption that
   the formed (NS/BH, RG) binary keeps the total angular momentum constant.
   According to Verbunt (1993), this timescale reads
   \begin{equation}
   \tau_{\rm sp}=\frac{1}{2(1-m_{\rm c}/M_{\rm RG})}\frac{M_{\rm RG}}{\dot{M}} 
                 \approx \frac{1}{2} \frac{M_{\rm RG}}{\dot{M}},
   \end{equation}
   where $\dot{M}$ is the rate of mass transfer onto the compact object.
   For the case of the RG as a donor in (NS/BH, RG)
   binary system, $\dot{M}$ can be
   estimated from Bhattacharya \& van den Heuvel (1991),
   \begin{equation}
   \dot{M}=1.4\times 10^{-2} M_1^{2.25}R_1^{0.75}~(M_{\odot}~{\rm yr^{-1}}),
   \end{equation}
   where $M_1=M_{\rm RG}/M_{\odot}$, and $R_1=R_{\rm RG}/R_{\odot}$.
   We thus obtain
   \begin{equation}
   \tau_{\rm sp}\approx 35.7 M_1^{-1.25}R_1^{-0.75}~({\rm yr}).
   \end{equation}
   From the above two equations we have the typical timescale for the 
   case of RG with
   mass $10M_{\odot}$, $\tau_{\rm sp}\sim$ 2 yr. This simply means
   that the new formed (NS/BH, RG) binary will reach the final stable
   configuration $D_f$ (listed in the Table 1) within 2 years, and
   become hard from the soft state very rapidly. It is thus viable
   to form hard binary in accretion disk. Furthermore, it has been
   shown in more detail by Magorrian \& Tremaine (1999) that the highest
   tidal disruption rate $10^{-4}$yr$^{-1}$ only take place in faint
   galaxies ($L<10^{10}L_{\odot}$). The tidal disruption of giant star
   could produce flare as often as every $10^5$yr. Therefore it is robust
   to believe that the present model of star-disk interaction works dominately
   over tidal capture by supermassive black hole hosting in the center of
   AGN/QSO in the active phase of AGN/QSO.

The detailed study of orbital changes due to the explosion of
supernovae can be found in Pols \& Marinus (1994), and Portegies Zwart 
\& Verbunt (1996). On the other hand, Verbunt(1993)
argued that this effect can be estimated by assuming that the
explosion is instantaneous, and the
positions and velocities of the stars are the same after the explosion as
before the explosion (Dewey \& Cordes 1987).
This implies that the distance $D_f$ between the compact star and the
core star before the explosion is the periastron distance after the
explosion and that the periastron velocity of the new orbit is the same
as the orbital velocity in the pre-supernova orbit. Therefore
the eccentricity is given by (Verbunt 1993)
\begin{equation}
e_{\rm ns} = {M_{\rm core} - m_{\rm ns} \over 2m_{\rm ns}}.
\end{equation}
The mass center of binary has obtained a speed $v_{\rm cm}$ because of the
mass loss and is given by
\begin{equation}
v_{\rm cm}=e_{\rm ns}v_1,
\end{equation}
where $v_1 = \sqrt{\frac{GM_{\rm core}}{D_f}} \sim 10^8
\left(\frac{M_{\rm core}}{4M\odot}\right)^{1/2}
\left(\frac{D_f}{1R_{\odot}}\right)^{1/2}$ cm s$^{-1}$
is the orbital velocity of the exploding component before explosion.
This velocity will make the (NS/BH, NS/BH) binary shooting up to a scale
height about $\sim 10h_{\rm ns}$. Furthermore, the orbit of the binary
is sufficiently small so it should be circularized by tidal interaction
and the radius $D_{\rm ns}$ of the circular orbit is given by
\begin{equation}
D_{\rm ns} = (1+e_{\rm ns})D_f.
\end{equation}
In table I, we list some possible values of $e_{\rm ns}$ and $D_{\rm ns}$ by
assuming $m_{\rm ns}=1.4M_{\odot}$. We can see that $e$ is larger than unity
when $M_{\rm RG}$ is larger than $15M_{\odot}$. This means that
after explosion the binary will be broken. For $m_{\rm ns}=3M_{\odot}$, 
we find that $e$ is always less than unity and D$_f$ is larger than R$_c$ 
even when $M_{\rm RG}$ equals 20$M_{\odot}$, in other words, there is no 
merger of massive black hole and helium core.

There are two interesting 
consequences of the new born (NS/BH, NS/BH) binary. First, the kick velocity
will carry the binary to leave the disk up to the 
distance about $10h_{\rm ns}\sim10^{17}cm$. Second, since
the separation of two compact stars in the (NS/BH, NS/BH) binary is 
$\sim D_{\rm ns}$, 
the gravitational radiation of the binary will make these two compact
stars merge in the time scale of (Peter 1964)
\begin{equation}
t_{\rm m}(D_{\rm ns},e)
  =\frac{12}{19}\left(\frac{D_{\rm ns}^4}{\beta c_1^4}\right) {\rm Int}(e),
\end{equation}
with
\begin{equation}
c_1=\frac{e^{\frac{12}{19}}}{(1-e^2)}
    \left(1+\frac{121}{304}e^2\right)^{\frac{870}{2299}},
\end{equation}
and
\begin{equation}
\beta=\frac{64}{5}\frac{G^3M_1M_2(M_1+M_2)}{c^5},
\end{equation}
and Int$(e)$ is the integral (see eq. 5.14 in Peter 1964). Therefore
the compact star merger rate should be about $\Gamma$. One should not
be surprised by the short time scale of merger since $t_m$
is proportional to $D_{\rm ns}^4$ while most of $D_{\rm ns}$ are less 
than solar radius, and $t_{\rm m}\propto (1-e^2)^{7/2}$ for larger 
eccentricity.
			       
\section{Discussions}

In this paper we suggest that if the metallicity of the metal-rich quasars 
result from the capture of 
the main sequence stars by the accretion disk surrounding the central
massive black hole of quasars and the captured stars 
can increase their masses by accretion 
so they can evolve off their main sequence branch and move toward the  
supernova stage which provide the observed metal abundance. Then the remnant 
stars are likely compact stars, i.e. neutron stars or black holes.  We have 
shown that the neutron star/black hole density in the vicinity of the disk 
will be high enough to a population of (NS/BH, RG) bnaries 
whic can eventually evolve to become either (i) (NS/BH, NS) binaries if 
the mass of RG is less than $15M_{\odot}$ and  merge in a time scale of 
a few million years or (ii) (NS/BH, Hc) binary.

In a wide range of cosmological GRB models the compact star merger plays 
an essential role. Since the main sequence star capture rate 
$\Gamma$ = 10$^{-3}$ yr$^{-1} \left(\frac{\rm NV/CIV}{0.01}\right)$ 
is proportional to 
the metal abundance ${\rm \frac{NV}{CIV}}$ and the compact object 
binary formation rate equals the 
main sequence star capture rate in our simple model. Therefore, the 
GBR burst rate $R_{\rm GRB}$ in AGNs or QSO can be estimated as 
\begin{equation}
R_{\rm GRB} \sim 10^{-3}\eta {\rm \frac{NV/CIV}{0.01} yr^{-1} QSO^{-1}},
\end{equation}
where $\eta$ is the beaming factor of the $\gamma$-rays which reduces the
observed probability of GRBs resulting from the compact star mergers
(e.g. $\eta$ is 0.1-0.01 for NS/NS merger, Ruffert, Janka, Schafer 1996;
Ruffert et al. 1997; Ruffert \& Janka 1998). Such estimated GRB 
rate in metal-rich quasars is much high than  that of ordinary
galaxies, e.g. the merger rate  due to the mass exchange
and gravitational radiation losses giving a conservative rate 
$\sim 10^{-6} yr^{-1}$ (Phinney 1991, Narayan et al 1992),  
or at a higher rate $10^{-(4\sim 5)}$ per year
(Tutukov \& Yungelson 1993, Portegies \& Verbunt 1996)
in light of the calculations of binary evolution.
Of course, the population of AGN/QSO is only 1\% of normal galaxies
which makes the total burst rate for these two classes of object comparable.
However the burst rate for individual AGN/QSO which has exceptional
high metal abundant could be very large. It has much higher possibility
that the not-yet-identified host of GRB may an AGN/QSO with very high
metallicity.

   Recently, Dokuchaev, Eroshenko \& Ozernoy (1998) propose the possibility of
   GRBs origin from the evolved galactic nuclei. In their model,
   GRBs result from  the coalescence of the compact object binary which
   are formed due to the dynamical evolution of cluster and dissipation
   of gravitational radiation. Carter (1992) proposed that
   GRBs are produced by the tidal disruption by the central supermassive
   black hole. Rees (1988) even pointed out that the rate of tidal
   disruption could be $10^{-4}$yr$^{-1}$ which is significantly
   lower than the capture rate [see eq.(6)] and formation rate of
   (NS/BH, NS/BH) binaries, particularly the whole star (unless
   giant star) will be swallowed
   by the supermassive black hole with mass $>10^8M_{\odot}$ since the Roche
   radius $R_t$ will lie within the horizon radius of the black hole. 
   Therefore it is believed in an ANG/QSO GRB rate due to the presently
   proposed process may dominate the tidal capture of star by the
   central black hole. 
One might argue that there is no AGN/QSO
at the locations of GRB970228 and GRB970508 which are detected by BeppoSAX
therefore it is unlikely that GRBs are related to this class of object.
However, in order to locate the accurate position
the BeppoSAX is restricted to long duration GRBs. Popham, Woosley \&
Fryer (1998) show that only (BH, Hc) merger can produce long duration burst,
other mergers [e.g. (NS/BH, NS/BH)] will not produce long duration burst.
   It is important to 
   note that the lifetime of AGN is typically of $10^8$ years,
   alternatively, the present paper proposed that at least some of
   GRBs may origin from the active nuclei due to the interaction between
   the star in cluster and accretion disk, roughly one percent of the
   observed GRBs.
In our model, only very low mass compact star and high mass RG binaries
can form (BH, Hc) mergers.
For a power law mass function, such binaries will be a very small fraction.
It is not too surprised not to identify any AGN/QSO with GRB970228 and
GRB970508 which are long duration bursts.
Interestingly there are two GRBs with small error box, in which AGN were found.
It was reported by Drinkwater et al (1997) that the possible X-ray
counterpart of GRB920501 is related with a Seyfert 1 galaxy at $z=0.315$.
Piro et al (1998) report that the first X-ray location of a $\gamma$-ray
burst by BeppoSAX (within $3^{\prime}$ radius) contains the  
quasar 4C 49 with $z=1.038$. Finally, we want to point out that the basic
energy mechanism of gamma-ray bursts in our model are the same as those
proposed by other authors, namely GRBs are resulting from the mergers of
compact binaries, except the formation rate of the binaries and the location
of Gamma-ray bursts. Since our model burst rate is proportional to the metal
abundance (cf. eq. 39), we shall predict that if the GRBs can be identified
with AGNs/QSOs, these AGNs/QSOs must have high metal abundance.

\acknowledgements 
The authors are grateful to the anonymous referee for his/her useful
comments on the surviving of formed binary in the dense star cluster
and other suggestions to improve the paper.
We thank Dr. R.Wijers for his useful comments and suggestions.
We are grateful to Prof. M.Veron for providing some information 
of quasars, Prof. D.N.C. Lin for discussion about the interaction 
between the main sequence stars and the accretion disk, Prof. R.E.
Taam for useful suggestions on the problems of spiral-in and the
capture of red giant stars, Prof. Z.W.Han for suggestions of the 
stellar evolution, Prof. Z.G.Dai for the discussion of 
gamma-ray bursts and Prof. M.A.Ruderman for suggestions and encouragement. 
This work is initiated in a Neutron Star Workshop in Aspen Center 
for Physics in 1995 and part of this work is supported by a RGC grant
of Hong Kong Government. JMW thanks the supports from the Climbing Plan of
the Ministry of Science and Technology of China and the National
Natural Science Foundation of China under grant No. 19803002.

\begin{center}
{Table 1 The calculation of binary formations$^*$}
\end{center}

\begin{center}
\begin{tabular}{ccccccccc} \hline \hline
$\mrg$&$M_{\rm c}$&$R_{\rm c}$&$D_i$&$D_f$&$e$&
$D_{\rm ns}$&$t_{\rm m}$(\rm yr)&Type\\ \hline  
10.0&2.155&0.349&296.021&0.621&0.270&0.788&0.804$\ttt 10^7$&(NS, NS/BH)\\
11.0&2.503&0.381&295.468&0.613&0.394&0.855&0.796$\ttt 10^7$&(NS, NS/BH)\\
12.0&2.869&0.414&295.003&0.607&0.525&0.925&0.635$\ttt 10^7$&(NS, NS/BH)\\
13.0&3.253&0.446&294.606&0.601&0.662&0.999&0.359$\ttt 10^7$&(NS, NS/BH)\\
14.0&3.655&0.479&294.264&0.596&0.805&1.077&0.987$\ttt 10^6$&(NS, NS/BH)\\
15.0&4.073&0.511&293.966&0.592&0.955&1.158&0.131$\ttt 10^5$&(NS, NS/BH)\\
16.0&4.507&0.543&293.704&0.589&$>1$ &.....&.....& .....\\
17.0&4.957&0.575&293.472&0.586&$>1$ &.....&.....& .....\\
18.0&5.423&0.607&293.265&0.584$<R_{\rm c}$&.....& ..... & .....&(Hc, BH)\\
19.0&5.903&0.638&293.079&0.582$<R_{\rm c}$&.....& ..... & .....&(Hc, BH)\\
20.0&6.398&0.670&292.911&0.580$<R_{\rm c}$&.....& ..... & .....&(Hc, BH)\\\hline
\end{tabular}
\end{center}
\begin{center}
$^*$ All parameters are in solar units except $e$ and $t_m(\rm yr)$.
 The mass of the compact star is taken to be 1.4M$_{\odot}$.
\end{center}
\end{document}